\documentstyle[preprint, aps, epsfig]{revtex}

\begin{document} 

\draft

\preprint{CINVESTAV--99--10} 
\tightenlines

\title{Testing flavor-changing neutral currents in the rare top quark
decays $t \to cV_iV_j$ \thanks{Work supported by CONACyT (M\'exico).}}

\author{J. L. D\'{\i}az-Cruz $^1$, M. A. P\'erez $^{2,3}$, G. 
Tavares-Velasco $^2$, and J.  J. Toscano $^3$}
\address{$^1$ Instituto de F\'{\i}sica, Universidad
Aut\'onoma de Puebla, Apartado Postal J-48, 72570, Puebla, M\'exico \\
$^2$ Departamento de F\'{\i}sica, CINVESTAV, Apartado Postal 14--740,
07000, M\'exico D. F., M\'exico \\ $^3$ Facultad de Ciencias F\'{\i}sico
Matem\'aticas, Universidad Aut\'onoma de Puebla, Apartado Postal 1152,
Puebla, M\'exico}

\maketitle 

\begin{abstract} 
We discuss the flavor-changing neutral currents (FCNC)  decays of the top
quark $t \to cV_iV_j$ ($V_i = \gamma ,Z ,g$) in the framework of the
Standard Model (SM) and in a two-Higgs doublet model (2HDM) with
tree-level FCNC couplings.  While in the SM the expected branching ratios
are extremely small, in the 2HDM they may be sizable, of order
$10^{-4}-10^{-5}$, and thus accessible at the CERN Large Hadron Collider.
We conclude with the interesting observation that the FCNC decay modes $t
\to cV_iV_j$ may not be equally suppressed as their corresponding decays
$t \to cV_i$ in this 2HDM.\\ 
\end{abstract}


Rare decay modes of the top quark have received considerable attention
recently because they might be a source of possible new physics effects to
be searched at future colliders.  Since the top quark has a mass of about
175 GeV \cite{abb}, the following rare processes may be kinematically
allowed: $t \to c \gamma$ \cite{diaz1}, $t \to cg, c Z$ \cite{grza}, $t
\to b W^+ Z$ \cite{jenk}, $t \to c W^+ W^-$ \cite{jenk,bar}, and $t \to
cZZ$ \cite{bar}.  Even though Standard Model (SM) predicts the branching
fractions for these rare decays to be unobservably small, it is believed
that the search for large signatures from flavor-changing neutral currents
(FCNC)  involving the top quark may serve as a unique test of the SM
\cite{bar,han}. The aim of this communication is to point out that the
FCNC transitions $t \to c \gamma \gamma$, $t \to c \gamma Z$ and $t \to c
g g$ also provide a detailed test of the SM and some of its extensions, e.
g., the two-Higgs doublet model (2HDM).

Just as it is the case with the FCNC one-vector boson decays $t \to c V_i$
($V_i=\gamma,g,Z$) \cite{diaz1,grza}, the two-vector boson decay modes $t
\to c V_i V_j$ are expected to be negligible in the SM because they are
induced at one-loop level and in addition the Glashow-Illiopoulos-Maiani
(GIM) mechanism suppresses effectively FCNC transitions involving virtual
down-type quarks.  However, in the most general version of the 2HDM (known
as Model III), where quarks are allowed to couple simultaneously to more
than one scalar doublet \cite{luke}, it is possible that important effects
will emerge in scalar FCNC couplings involving quarks of the second and
third generations. Unlike earlier versions of the 2HDM, in Model III no ad
hoc symmetries are invoked to eliminate tree-level scalar FCNC couplings,
but instead a more realistic pattern for the Yukawa matrices is imposed
and constraints on the scalar FCNC are derived from phenomenology
\cite{attw1}.  The tree-level scalar FCNC interactions are given by

\begin{equation} 
\label{lag}
{\cal{L}}_{Y,FCNC}^{III} = \xi_{ij} \sin{\alpha} \bar{f_i} f_j h^0 +  
\xi_{ij} \cos{ \alpha} \bar{f_i} f_j H^0 + 
\xi_{ij} \cos{\alpha} \bar{f_i} \gamma^5 f_j A^0 + 
h.c.,
\end{equation} 

\noindent where we are using the Higgs mass-eigenstate basis with the
light and heavy CP-even Higgs bosons $h^0$ and $H^0$, respectively, and
the CP-odd Higgs boson $A^0$, $\alpha$ denotes the mixing angle, and
$\xi_{ij}$ corresponds to the off-diagonal Yukawa couplings. It is
convenient to use the parametrization introduced in \cite{chengsher}: 
$\xi_{ij}= \lambda_{ij} \frac{\sqrt{m_i m_j}}{v}$, where the mass factor
gives the strength of the interaction while the dimensionless parameters
$\lambda_{ij}$ are fixed by low energy experiments. Although there are
strong bounds on couplings involving light quarks, no stringent bounds
exist for $\lambda_{tc}$, and it is feasible a less suppressed strength
for the interaction $tc\phi_k^0$, with $\phi_k^0$ any of the three
physical Higgses of Model III.

In the SM and Model III the FCNC decays $t \to c V_i V_j$ may proceed
through the resonant diagram shown in Fig. 1, where $\phi_k^{0 \ *}$
represents the exchanged Higgs, and the dots denote the couplings of this
scalar with the particles in the initial and final states. The amplitude
for this diagram can be written in general as

\begin{equation} 
\label{mtcvv} 
{\cal{M}}(t \to c V_i V_j)= \sum_k P_{\phi_k^0}(s) \ {\cal{M}}(t \to c
\phi^{0 \ *}_k) \ {\cal{M}}(\phi^{0 \ *}_k \to V_i V_j),
\end{equation}

\noindent where $P_{\phi^0_k}(s)$ is the resonant propagator of the
exchanged Higgs, which transfers a squared 4-momentum $s=(k_1+k_2)^2$,
which in turn is restricted by kinematics to lie in the range
$(m_{V_i}+m_{V_j}) \leq \sqrt{s} \leq (m_t-m_c)$. The partial amplitudes
in Eq.  (\ref{mtcvv})  can be expressed in terms of the $\phi^0_k$
couplings

\begin{equation} 
\label{mtch} 
{\cal{M}}(t \to c \phi^{0 \ *}_k)=\bar{u}(q) \ \Gamma_{t c \phi^0_k} \
v(p),
\end{equation}

\begin{equation} 
\label{mhvv} 
{\cal{M}}(\phi^{0 \ *}_k \to V_i V_j)= i \Gamma_{k \ \mu \nu}^{V_i V_j}\
\epsilon^\mu (k_1,\lambda_1) \epsilon^\nu (k_2,\lambda_2),
\end{equation}

\noindent where $\Gamma_{t c \phi^0_k}$ characterizes the coupling $t c
\phi^0_k$, which in the SM arises at one-loop level but in Model III it is
a tree-level effect, whereas $\Gamma_{k \ \mu \nu}^{V_i V_j}$
characterizes the one-loop level coupling $\phi^0_k V_i V_j$ \cite{gun}. 
In the case of the channel $t \to c g g$, Eq. (\ref{mhvv}) must be
modified to account for the $\mathrm{SU(3)}$ gauge structure of gluons. 
The decay width $\Gamma(t \to c V_i V_j)$ can be obtained in the usual way
and it is given by

\begin{equation} 
\label{width} 
\Gamma (t \to c V_i V_j)=\frac{1}{256 \pi^3 m_t}
\int^{(mt-mc)^2}_{(m_{V_i}+m_{V_j})^2} \lambda (m_t,m_c,\sqrt{s}) 
\lambda(\sqrt{s},m_{V_i},m_{V_j}) \ |{\cal{M}}(t \to c V_i V_j) |^2 ds,
\end{equation}

\noindent where $\lambda(x,y,z)$ is a phase space factor. 

To realize how the top quark decays $t \to c V_i V_j$ get enhanced in
Model III, it is necessary to examine the branching ratios in the SM,
where only the FCNC two-vector boson decay $t \to c W^+ W^-$ occurs at
tree-level with a branching ratio of order $10 ^{-14}-10^{-12}$
\cite{jenk}. As far as the decay modes $t \to c \gamma \gamma$, $ t \to c
\gamma Z$ and $t \to c gg$ are concerned, in addition to the one-loop
Feynman graphs, the resonant diagram shown in Fig. 1 contributes to the
decay amplitude.  In this case the SM Higgs $\phi^0_{\mathrm{SM}}$ is the
mediator and the coupling $tc\phi^0_{\mathrm{SM}}$ is induced at one-loop
level.  The contribution of this diagram seems to be of order $g^6$, but a
careful analysis shows that on-resonance it becomes of order $g^4$, being
the dominant contribution for light bosons and competing with the one-loop
contributions \cite{diaz2}. The SM coupling $tc\phi^0_{\mathrm{SM}}$ can
be expressed as

\begin{equation}
\label{tchver}
\Gamma_{tc\phi^0_{\mathrm{SM}}}=\sum_{k=d,s,b}V_{t k} V^*_{k c}
\left(A_L(s,m_k^2) P_L +
A_R(s,m_k^2)
P_R \right),
\end{equation}

\noindent where $P_L$ and $P_R$ are, respectively, the left- and
right-handed chirality projectors, whereas $A_L$ and $A_R$ are given in
terms of scalar integrals. We have calculated
$\Gamma_{tc\phi^0_{\mathrm{SM}}}$ and our result \cite{diaz2} agrees with
the one presented in \cite{eil1} and later reproduced in \cite{eil2,petr}
for the SM decay $t \to c \phi^0_{\mathrm{SM}}$. This coupling has also
been studied in the Minimal Supersymmetric Standard Model \cite{yang}. In
the zero charm quark mass limit the function $A_R$ vanishes and can be
neglected in the calculation. From Eqs. (\ref{mtcvv})-(\ref{tchver}) it is
straightforward to obtain, after dividing by the main top quark width
$\Gamma(t \to b W)$, the SM values $B(t \to c \gamma \gamma)  \simeq
10^{-15}-10^{-16}$, $B(t \to c \gamma Z)\simeq 10^{-15}-10^{-16}$, and
$B(t \to c g g) \simeq 10^{-14}-10^{-15}$ for $m_Z \leq
m_{\phi^0_{\mathrm{SM}}} \leq m_t$. 

In the framework of Model III we have studied the decays $t \to c \gamma
\gamma$, $t \to c \gamma Z$, $t \to c g g$, and for completeness also the
channel $t \to c W^- W^+$, for which we agree with the results found in
\cite{bar}. We will not discuss the mode $t \to c Z Z$ because it is
highly restricted by phase space. Since the most relevant phenomenological
tree-level FCNC effects are due to the light scalar $h^0$ \cite{attw1},
the heavy ones $H^0$ and $A^0$ are unimportant for our present discussion.
In fact one can assign to them an arbitrary large mass, namely $750$ TeV.
The same value was assigned to the charged Higgs mass $m_{H^\pm}$, which
enters in the expressions for the one-loop induced coupling $\phi^0_k V_i
V_j$, though we found that the results for the decay widths are not
sensitive to this choice. For simplicity we have chosen $\lambda_{tc}$ to
be real, thus the branching ratios obtained from Eq. (\ref{width}), after
dividing by $\Gamma(t \to bW)$, turn out to be proportional to the factor
$f_{\lambda}=\lambda_{tc} \sin \alpha \cos\alpha$. Plots for the branching
ratios scaled by the factor $f_{\lambda}$ are shown in Fig.  2 as a
function of the Higgs mass $m_{h^0}$. This graph teaches us that the
branching ratios reach their most significant enhancement in the resonance
region $(m_{V_i}+m_{V_j})  \leq m_{h^0} \leq (m_t-m_c)$, where the
channels $t \to c \gamma \gamma$ and $t \to c \gamma Z$ may have branching
fractions as high as $10^{-5}$-$10^{-4}$, while those of the modes $t \to
c W^+ W^-$ and $t \to c g g$ may reach values up to $10^{-4}$. The
enhancement is due to the combined effect of the tree-level FCNC and the
resonance of the Higgs boson.  On-resonance, the decay $t \to c V_i V_j$
proceeds in two stages: first the top quark decays as $t \to c h^0$, with
the on-shell Higgs decaying then as $h^0 \to V_i V_j$. This contrasts with
the pure three-body decay that occurs outside the resonance region.
Although in Model III the decay $t \to c W^+ W^-$ is induced at
tree-level, it is highly suppressed by phase space, and its width is only
one order of magnitude greater those for the decays $t \to c \gamma
\gamma$ and $t \to c \gamma Z$.

The possibility of detecting FCNC top quark transitions at the Tevatron
has been previously investigated, for instance through the reaction $c+g
\to t+A^0$ \cite{taiwaness}. Here we will discuss the mode $t\to
c+\gamma\gamma$ as a representative case of new signatures associated with
FCNC top quark decays. This reaction turn out to be very clean when it is
attempted to separate the signal from background. Let us consider the
situation at the Tevatron, where it will be possible a production of top
quark pairs through quark and gluon fusion with a cross-section of
$5\times 10^3$ fb \cite{topyuan}. If the top quark decays into
$c+\gamma\gamma$ with a branching ratio of order $10^{-5}$, an integrated
luminosity of $10^2$ fb$^{-1}$ would be needed in order to produce at
least one event, which exceeds the planned luminosity (of order 1
fb$^{-1}$).  Then it seems hopeless to detect the FCNC two-photon top
quark decay at the Tevatron.  Nevertheless it remains the possibility of
detecting the Higgs decay $A^0 \to t\bar{c}$, which may have a branching
ratio of order unity for $m_{A^0} \geq m_t+m_c$. At the Tevatron the
cross-section for the reaction $gg \to A^0 \to t\bar{c}$ is about $10^3$
fb, and the same is true for the cross-section of the main background,
which is expected to come from the charged-current reaction $q+b \to c+t$.
Thus it seems feasible that an adequate implementation of cuts will allow
to identify the signal.

On the other hand, the situation looks more promising at the CERN Large
Hadron Collider (LHC), where the cross-section to produce top quark pairs
will be $4.3 \times 10^5$ fb, and a branching ratio of order $10^{-5}$
will reduce this cross-section down to 4.3 fb. We will use a conservative
efficiency (0.05) for the production of the $t+c +\gamma +\gamma$ events
at the LHC. We obtained this efficiency from the detailed study of T. Han
{\it{et al.}} \cite{han} on the single top quark production via FCNC
couplings at hadron colliders. With an integrated luminosity of 100
fb$^-1$ we then get only 21 $t +c +\gamma +\gamma$ events per year at the
LHC. The main background to this signal comes from the production of $qb
\to t+c+\gamma+\gamma$ and $q\bar{q} \to t +b +\gamma +\gamma$ events,
where the final $b$ may be misidentified as $c$. The analysis in
\cite{han} for the single top quark production can be used also to show
that, after the inclusion of the suppression factor $(\frac{4 \alpha}{9
\pi})^2$ to account for the two emitted photons, these background
processes will have less than one event at the LHC and thus will make
feasible the detection of our signal. Even more, our $t + c +\gamma
+\gamma$ signal is so distinctive that its detection should be difficult
to miss at the LHC: two highly monochromatic photons with an invariant
mass peaked at $m_{\phi^0_k}$ plus a single quark and a jet. A more
systematic study of all possible backgrounds is, of course, beyond the
scope of the present rapid communication.

Let us consider another implications of our results. If we assume
$\lambda_{tc}$ to be of order unity, the branching ratios for $t \to c V_i
V_j$ may reach values up to $10^{-4}$ in the 2HDM, while the branching
ratios for the decays $t \to c V_i$ ($V_i=\gamma, Z, g$) could be of order
$10^{-5}-10^{-11}$, depending on the Higgs boson masses and the value of
$\tan{\beta}=\case{\nu_2}{\nu_1}$, with $\nu_{1,2}$ the vacuum expectation
values of the two Higgs doublets \cite{eil2}. Thus, we find that for a
light Higgs the three-body decay modes $t \to c V_iV_j$ may not be equally
suppressed as the two-body decays $t \to c V_i$ in the 2HDM.  Our result
represents an explicit realization of the scenario previously conjectured
in studies of the rare processes $b \to s \gamma \gamma$ \cite{lin}, $\mu
\to e \gamma \gamma$ \cite{bowm} and $\nu^{'} \to \nu \gamma \gamma$
\cite{niev}, where the corresponding branching ratios get enhanced in
comparison with those of the one-photon processes.

A large two-photon FCNC transition may also have implications for the
decay $B^0_s \to \gamma \gamma$, which can occur through an annihilation
graph mediated by the Higgs boson $A^0$. A simple PCAC evaluation shows
that the corresponding branching ratio is of order $10^{-7}$, which is
comparable to the SM result. In order to obtain further conclusions one
needs to include QCD corrections, as has been done recently for the SM and
virtual contributions of the charged Higgs \cite{aliev}.

To close this communication we would like to stress the relevance of our
results.  Even though the SM is a well established theory supported by a
plethora of experimental data, it leaves us with some unanswered
questions. In particular, the scalar sector with its elusive Higgs boson
remains as the SM most puzzling piece. Supersymmetric Grand Unified
Theories allow for a relatively light Higgs boson in the range $100-200$
GeV \cite{car}, and it is possible that in the next years it will occur
the milestone discovery of a light Higgs boson.  Meanwhile there is an
incessant search for any evidence of new physics that could enlighten the
road to a more comprehensive theory of elementary interactions. In this
context, the results presented in this rapid communication open up the
possibility of looking for signals from physics beyond the SM in the FCNC
top quark decays $t \to c \gamma \gamma$, $t \to c \gamma Z$, $t \to c W^+
W^-$, and $t \to c g g$.

\begin{figure} 
\label{fig1}
\begin{center}
\epsfig{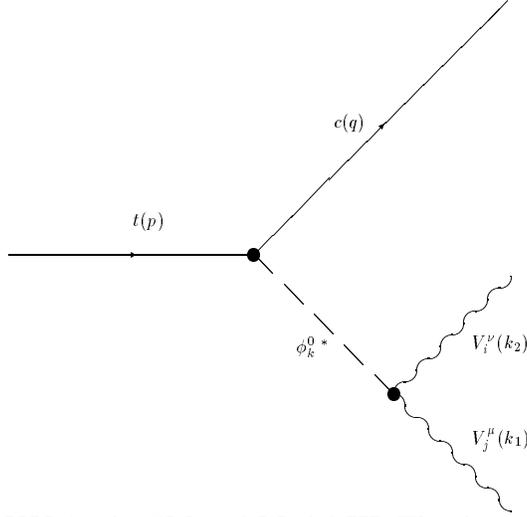}
\caption{The decay $t \to c V_i V_j$ in the SM and Model III. The dots
denote couplings that can be induced at tree or one-loop level.} 
\end{center}
\end{figure}

\begin{figure} 
\label{fig2} 
\begin{center}
\epsfig{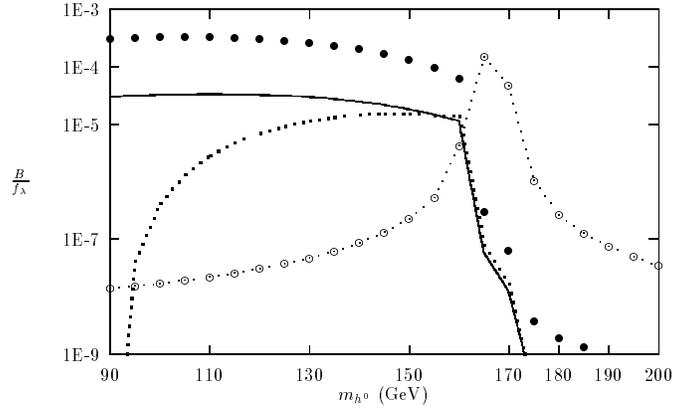} 
\end{center} 
\caption{Scaled branching ratios for $t \to c V_i V_j$ in Model III: $t
\to c \gamma \gamma$ (solid line), $t \to c \gamma Z$ (points), $t \to c
W^+ W^-$ (hollow circles), and $t \to c g g$ (full circles). A value of
750 TeV is used for the masses $m_{A^0}$, $m_{H^0}$ and $m_{H^\pm}$.}
\end{figure}


\begin{references} 

\bibitem{abb} B. Abbott {\it{et al.}}, D0 Coll., Phys. Rev. D {\bf{58}},
052001 (1998). 

\bibitem{diaz1} J.L. D\'{\i}az-Cruz, R. Martinez, M.A. P\'erez, and A.
Rosado, Phys. Rev. D {\bf{41}}, 891 (1990).

\bibitem{grza} B. Grzadkowski, J.F. Gunion, and P. Krawczyk, Phys. Lett. 
{\bf{B268}}, 106 (1991); E. Eilam, J.L. Hewett, and A. Soni, Phys. Rev. D
{\bf{44}}, 1473 (1991); M. Luke and M.J. Savage, Phys Lett. {\bf{B307}},
387 (1993); G.  Couture, C.  Hamzauni, and K. K\"onig, Phys. Rev. D
{\bf{52}}, 1713 (1995). 

\bibitem{jenk} E. Jenkins, Phys. Rev. D {\bf{52}}, 1713 (1995).

\bibitem{bar} S. Bar-Shalom, G. Eilam, A. Soni, and J. Wudka, Phys. Rev. 
Lett.  {\bf{19}}, 1217 (1997); Phys. Rev. D {\bf{57}}, 2957 (1998). 

\bibitem{han} T. Han, K. Whisnant, B. -L. Young, and X. Zhang, Phys. Rev D
{\bf{55}}, 7241 (1997). 

\bibitem{luke} M. Luke and M. J. Savage, Phys. Lett {\bf{B307}}, 387
(1993); T. P.  Cheng, M. Shar, Phys. Rev. D {\bf{35}}, 3484 (1987). 

\bibitem{attw1} For a review of phenomenology of Model III see: M. Sher,
hep-ph/9809590; D. Atwood, L. Reina, and A. Soni, hep-ph/9612388; D. 
Atwood, L. Reina, and A. Soni, Phys. Rev. D {\bf{55}}, 3156 (1997);  J.L. 
D\'{\i}az-Cruz, Phys. Rev. D {\bf{51}}, 5263 (1995).

\bibitem{chengsher} T.P. Cheng and M. Sher, Phys. Rev. D {\bf{35}},3484
(1987); A. Antaramian, L. Hall and A. Rasin, Phys. Rev. Lett. {\bf{69}},
1871 (1992). 

\bibitem{gun} J. Gunion, H. Haber, G. Kane, and S. Dawson, {\it{The Higgs
Hunter's Guide}}, Addison Wesley, 1990. 

\bibitem{diaz2} J. L. D\'{\i}az-Cruz, M. A. P\'erez, G. Tavares-Velasco,
and J. J. Toscano, in preparation.

\bibitem{eil1} G. Eilam, B. Haeri, and A.Soni, Phys. Rev. D {\bf 41}, 875
(1991).

\bibitem{eil2} G. Eilam, J. L. Hewett, and A. Soni, Phys. Rev. D
{\bf{44}}, 1473 (1991); {\it{ibid}} D {\bf{59}} 039901 (E), 1999. 

\bibitem{petr} B. Mele, S. Petrarca and A. Soddu, Phys. Lett. {\bf B435},
401 (1998).

\bibitem{yang} J. Yang and C. Li, Phys. Rev. D {\bf{49}}, 3412 (1994);
{\it{ibid}} D {\bf{51}} 3974 (E), 1995.

\bibitem{taiwaness} W. Hou, G. Lin, C. Ma, and C.P. Yuan,
Phys. Lett. {\bf B409}, 344 (1997)

\bibitem{topyuan} C.P. Yuan in {\it{Lectures on top quark physics}},
Procc. of the VI Mexican School of Particles and Fields, edited by J. C. 
D'Olivo, M. Moreno, and M.A. P\'erez, World Scientific, Singapore, 1995. 

\bibitem{han} T. Han {\it{et al.}}, Phys. Rev. D 58, 073008 (1998).

\bibitem{lin} G. -L. Lin, J. Liu, and Y. -P. Yao, Phys. Rev. Lett. 
{\bf{64}},1498 (1990); Phys. Rev. D {\bf{42}}, 2314 (1990). 

\bibitem{bowm} J. D. Bowman, T. P. Cheng. L. F. Li, and H. S. Matis, Phys. 
Rev. Lett. {\bf{41}}, 442 (1978); J. Dreitlein, and H. Primakoff, Phys. 
Rev.  {\bf{126}}, 375 (1962).

\bibitem{niev} J. F. Nieves, Phys. Rev. D {\bf{28}}, 1664 (1983).

\bibitem{aliev} T.M. Aliev and E.O. Itan, Phys. Rev. D {\bf{58}}, 095014
(1998); C. V. Chang, G.-L. Lin, and Y. -P. Yao, hep-ph/9705345.

\bibitem{car} M. Carena, {\it{et al.}} in {\it{Higgs Physics at LEP2}},
CERN 96-01, edited by G. Altarelli, T. Sj\"ostrand, and F Zwiner.

\end{references}
\end{document}